\documentclass{article}

\usepackage{amssymb}
\usepackage{slashbox}
\usepackage{amsmath}
\usepackage{float}
\usepackage[backend=biber]{biblatex}
\usepackage{authblk}
\usepackage{graphicx}

\newtheorem{theorem}{Theorem}[section]
\newtheorem{proposition}{Proposition}
\newtheorem{definition}{Definition}

\newtheorem{corollary}{Corollary}[section]

\newtheorem{example}{Example}
\newtheorem{remark}{Remark}

\newcommand{\G}{\Gamma}

\title{The search of Type I codes}

\author[1]{Carolin Hannusch}
\author[2]{Roland S. Major}

\affil[1]{Faculty of Informatics, University of Debrecen\\ hannusch.carolin@inf.unideb.hu }
\affil[2]{Faculty of Informatics, University of Debrecen\\ major.sandor@inf.unideb.hu}

\begin{document}
	\maketitle
	
	\begin{abstract}
		A self-dual binary linear code is called Type I code if it has singly-even codewords, i.e.~it has codewords with weight divisible by $2.$ The purpose of this paper is to investigate interesting properties of Type I codes of different lengths. Further, we build up a computer-based code-searching program based on our knowledge about Type I codes. Some computation results achieved by this program are given.\\ 
		Keywords: Type I codes; self-dual linear codes\\
		AMS Subject Classification: 94B05
	\end{abstract}
	
	\section{Preliminaries and Motivation}\label{sec:1}

	A linear code of length $n$ is a subspace of the $n$-dimensional vectorspace over a field $\mathbb{F}$. Throughout our paper, all linear codes are binary codes, i.e.~$\mathbb{F}=\mathbb{F}_2.$ Let $C$ be a linear code of length $n$, dimension $k$ with minimum distance $d.$ Then we say that $C$ is an $(n,k,d)$-code. Any two codewords $x$ and $y$ are orthogonal if their inner product is $0.$ Each linear code has a dual code $C^{\bot}=\{c\in \mathbb{F}_2^n\mid c \bot x \mbox{  } \forall x\in C\}.$ If $C=C^{\bot}$, then $C$ is called self-dual. In this case we have $k=\frac{n}{2}.$ It is well known (see e.g.~Theorem 2.19 in \cite{lindell2010introduction}) that each code $C$ is equivalent to a linear code generated by a generator matrix $G$ of the form 
	
	$$G=\left(\begin{array}{ccc}
		I_{k} & | & A
	\end{array}\right),$$
	
	where $I_{k}$ denotes the identity matrix of dimension $k$ and $A \in \mathcal{M}_{k\times k}(\mathbb{F}_2).$ If $C$ is self-dual and it has codewords of weight divisible by $2,$ then $C$ is called Type I code. Otherwise (if $C$ has only codewords of weight divisible by $4$) $C$ is called Type II code. By \cite{rains1998shadow} we have the same upper bound for the minimum distance of self-dual Type I and Type II codes, namely $d\leq 4\lfloor\frac{n}{24}\rfloor+4$ and $d\leq 4\lfloor\frac{n}{24}\rfloor+6$ if $n\equiv 22 \mod 24.$
	
	Type I codes were investigated in \cite{bouyuklieva2017nonexistence}, \cite{bouyuklieva2012singly} and \cite{harada2010new}.
	The investigation of Type I codes is interesting, since several open problems can be found in literature considering self-dual Type I codes (e.g.~\cite{dougherty2015open}, \cite{joyner2011selected}). For example, the existence of a Type II self-dual $(72,36,16)$ code is still an open question and it is equivalent to the existence of a Type I self-dual $(70,35,14)$-code (\cite{dougherty2015open}.) Furthermore, the existence of a Type I self-dual $(56,28,12)$-code is  also still an open question. 
	
	This paper is organized as follows. In Section \ref{sec:2} we summarize and introduce the main theoretical background for our code searching program. In Section \ref{sec:3} we investigate the relation of Type I codes to its neighbors. In Section \ref{sec:4} we describe the code searching package itself. Afterwards, in Section \ref{sec:5} some computational results are given. Finally, in Section \ref{sec:6} we draw our conclusion so far.
	
	\section{Properties of Type I codes}\label{sec:2}
	
	\subsection{General properties}\label{general}
	
	\begin{proposition}
		Let $n\equiv 0 \mod 4$ and $C$ a self-dual binary $(n,k,d)$-code. If $G$ is a generator matrix of $C$ in standard form, then the number of rows in $G$ with singly-even weight is even.
	\end{proposition}
	
	\textbf{Proof. }
		We denote the generator matrix by $G=\left( \begin{array}{c}
			g_1\\
			\vdots\\
			g_k	
		\end{array}\right).$ The proposition follows by the fact that $\textbf{1}\in C$ and $\textbf{1}=\sum_{i=1}^k g_i.$  Since $w(c_1+c_2)=w(c_1)+w(c_2)-2\mu(c_1,c_2),$ we have $w(c_1+c_2)\equiv 0 \mod 4 \Leftrightarrow w(c_1)\equiv w(c_2) \mod 4.$
	$\hfill\square$
	
	\begin{corollary}
		Let $G$ be a generator matrix of a Type I code in standard form. Then $G$ has at least two singly-even rows.
	\end{corollary}
	
	\subsection{Graph of a generator matrix}\label{graph}
	
	\begin{definition}\label{tree}
		Let $A$ be a binary matrix with rows $a_1,\ldots, a_k$ and $w(a_1)\geq \ldots \geq w(a_k).$ Further the $1$-s in row $a_1$ are in consecutive order on the lefthand side, and the $0$-s are in consecutive order on the righthand side, i.e.~there exists a number $1\leq t\leq k,$ such that
		$$a_1=(\underbrace{1 \ldots 1}_{t}\underbrace{0\ldots 0}_{k-t}).$$
		
		Let $\G$ be a tree graph corresponding to $A$ defined in the following way: It's root is the set $\{1,\ldots, k\}$. The root has two child nodes: One is the subset $S_1 \subset  \{1,\ldots, k\}$ containing those positions which are $1$ in $a_1.$ Thus $S_1=\{1,\ldots,t\}.$ The other one is the subset $S_2 \subset  \{1,\ldots, k\}$ containing those positions which are $0$ in $a_1.$ Thus $S_2=\{t+1,\ldots,k\}.$ Each of those nodes has again two child nodes, defined in the same way. We repeat this step for all nodes and branches until we get $k$ levels. 	
	\end{definition}
	
	\begin{example}
		Let $$A=\left(\begin{array}{cccccc}
			1 & 1 & 1 & 1 & 1 & 0\\
			1 & 1 & 1 & 0 & 0 & 1\\
			1 & 0 & 0 & 1 & 0 & 1\\
			1 & 0 & 0 & 0 & 1 & 0\\
			0 & 0 & 0 & 1 & 1 & 0\\
			0 & 0 & 0 & 1 & 0 & 1
		\end{array}\right)$$
		
		Then the graph $\G$ corresponding to $A$ is:
		\begin{center}
			\includegraphics[scale=0.5]{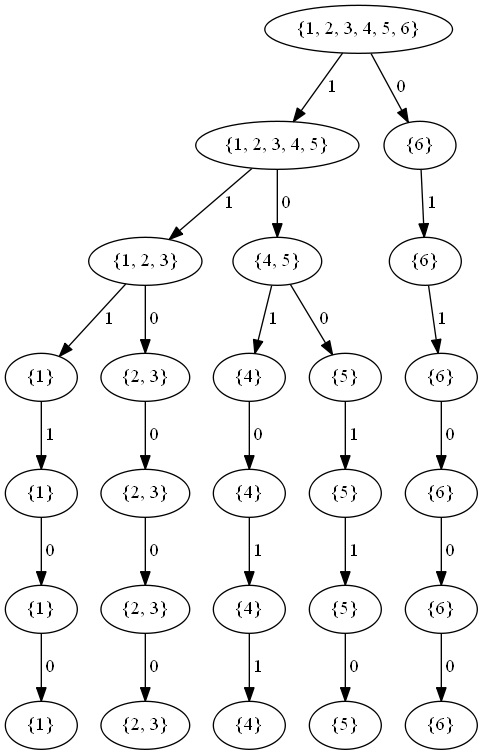}
		\end{center}
	\end{example}
	
	\begin{remark}
		The graph $\G$ has at most $k$ leaf nodes.
	\end{remark}
	
	\begin{proposition}
		Let $A$ be a binary matrix with $k$ rows and let $\G$ be its corresponding graph defined in Definition \ref{tree} . $A$ exists if and only if $\G$ is connected, it has $k+1$ levels and the union of the leaf nodes is the set $\{1,\ldots,k\}.$
	\end{proposition}
	
	\textbf{Proof. } Let us denote the nodes of $\G$ by $S_0, S_1, S_2, \ldots, S_m,$ where $S_0 =\{1,\ldots, k\}$ and $S_i\subseteq S_0$ for each $i\in\{1,\ldots,m\}$ and $m$ is an arbitrary integer. By definition, $\G$ has $k+1$ levels. We use induction on $k.$
		If $k=1,$ then $\G$ has 2 levels and the matrix $A_{1\times k}$ can be constructed if $S_1\cup S_2 = S_0.$ 
		We assume that the statement is true for $k=f.$ 
		Now, let $k=f+1.$ Then the $f+1^{th}$ row of $A$ can be constructed by the $f+1^{th}$ level in $\G.$ Any coordinate of this row has an entry only if it is contained in any leaf node. Thus if the union of the leaf nodes is $\{1,\ldots,k\},$ then the $f+1^{th}$ row of $A$ can be constructed. Otherwise, it cannot be constructed. 
	$\hfill\square$
	
	\subsection{The $\mu$ table}\label{mu}
	
	Let $C$ be a binary self-dual $(n,k,d)$ code. For all $x,y \in C$ we have 
	$w(x+y)=w(x)+w(y)-2\mu(x,y),$ where $\mu(x,y)$ denotes the number of coordinates which are $1$ in both, $x$ and $y.$ Since $d\leq w(x+y)\leq n-d,$ we get borders for $\mu(x,y).$ For example, we search for a singly-even $(56,28,12)$ Type I code. If it exists, then it has a generator matrix $G=(I_{28} \mbox{  } A).$ Let $r_1$ and $r_2$ be two arbitrary rows of $A.$ We summarize the possible values for $\mu(r_1,r_2)$ in the following table.

	\begin{table}[htbp]\caption{Possible values of $\mu(r_1,r_2)$}
		\begin{tabular}{ *{10}{|c}|} \hline
			\backslashbox[10mm]{$w(r_1$)}{$w(r_2$)}	&\makebox[3em]{27}&\makebox[3em]{25}&\makebox[3em]{23}&\makebox[3em]{21}&\makebox[3em]{19}\\\hline\hline
			27 & - & - & - & - & $\{18\}$  \\\hline
			25 & - & - & - & $\{18\}$ & $\{16\}$\\\hline
			23 & - & - &  $\{18\}$ & $\{16\}$ & $\{14,16\}$\\\hline
			21 & - &  $\{18\}$ & $\{16\}$ & $\{14,16\}$ &$\{12,14\}$\\\hline
			19 & $\{18\}$ & $\{16\}$ & $\{14,16\}$ &$\{12,14\}$ &$\{10,12,14\}$\\\hline
			17  & $\{16\}$ & $\{14,16\}$ &$\{12,14\}$ &$\{10,12,14\}$ & $\{8,10,12\}$\\\hline
			15  & $\{14\}$ &$\{12,14\}$ &$\{10,12,14\}$ & $\{8,10,12\}$ & $\{6,8,10,12\}$\\\hline
			13  & $\{12\}$ &$\{10,12\}$ & $\{8,10,12\}$ & $\{6,8,10,12\}$& $\{4,6,8,10\}$ \\\hline
			11 & $\{10\}$ & $\{8,10\}$ & $\{6,8,10\}$ & $\{4,6,8,10\}$ & $\{2,4,6,8,10\}$ \\\hline
			
		\end{tabular}
		
		\begin{tabular}{ *{10}{|c}|} \hline
			\backslashbox[10mm]{$w(r_1$)}{$w(r_2$)}	&\makebox[3em]{17}&\makebox[3em]{15}&\makebox[3em]{13}&\makebox[3em]{11}\\\hline\hline
			27 & $\{16\}$ & $\{14\}$ &$\{12\}$ &$\{10\}$\\\hline
			25 & $\{14,16\}$ &$\{12,14\}$ &$\{10,12\}$ & $\{8,10\}$\\\hline
			23 &$\{12,14\}$ &$\{10,12,14\}$ & $\{8,10,12\}$ & $\{6,8,10\}$\\\hline
			21 &$\{10,12,14\}$ & $\{8,10,12\}$ & $\{6,8,10,12\}$ & $\{4,6,8,10\}$\\\hline
			19 & $\{8,10,12\}$ & $\{6,8,10,12\}$ & $\{4,6,8,10\}$ & $\{2,4,6,8,10\}$\\\hline
			17 & $\{6,8,10,12\}$ & $\{4,6,8,10\}$ & $\{2,4,6,8,10\}$ & $\{0,2,4,6,8\}$\\\hline
			15 & $\{4,6,8,10\}$ & $\{2,4,6,8,10\}$ & $\{0,2,4,6,8\}$ & $\{0,2,4,6,8\}$\\\hline
			13 & $\{2,4,6,8,10\}$ & $\{0,2,4,6,8\}$ & $\{0,2,4,6,8\}$ & $\{0,2,4,6\}$\\\hline
			11 & $\{0,2,4,6,8\}$ & $\{0,2,4,6,8\}$ & $\{0,2,4,6\}$& $\{0,2,4,6\}$\\\hline
			
		\end{tabular}
	\end{table}

	\begin{proposition}\label{chain}
		Let $C$ be a singly-even self-dual code of length divisible by $4.$ Then there exists a chain of subcodes $$C_1=\left\langle \textbf{1}\right\rangle \subset C_2 \subset \ldots \subset C_{k-1}\subset C,$$ where $C_i$ is doubly-even self-orthogonal for $i=1,\ldots,k-1.$
	\end{proposition} 
	
	\textbf{Proof. } We denote the length of $C$ by $n$ and its dimension by $k.$ On the one hand $n\equiv 0 \mod 4,$ which implies $\textbf{1}$ is a doubly-even codeword of $C.$ On the other hand, we know that each singly-even code has a maximal doubly-even subcode of 1-codimension. The existence of the subcodes between $1$ and $C_{k-1}$ follows from here.
	$\hfill\square$
	
	\section{Neighbors}\label{sec:3}
	
	Pless and Brualdi \cite{brualdi1991weight} introduced the concept of neighbors within  self-dual codes. In this section we investigate the neighbors of Type I codes.
	
	\begin{definition}
		Two self-dual codes of length $n$ are called neighbors, provided their intersection is a code of dimension $\frac{n}{2}-1.$
	\end{definition}
	
	\begin{theorem}\label{eitheror}
		Let $C$ be a self-dual code of Type I. Then $C$ has either two doubly-even neighbors or two singly-even neighbors.
	\end{theorem}
	
	\textbf{Proof. }
		We know that $C$ has a maximal doubly-even subcode $C_{max}\subset C.$ Then there exists $\gamma_1\in C_{max}^{\bot}$ such that $C=\langle C_{max},\gamma_1\rangle,$ i.e.~$w(\gamma_1)\equiv 2 \mod 4.$ Let $\gamma_2\in C_{max}^{\bot}.$ Since $C$ is self-dual we have $\gamma_1\not\perp \gamma_2$ and $C_1=\langle C_{max},\gamma_2\rangle$ and $C_2=\langle C_{max},\gamma_1+\gamma_2\rangle$ are neighbors of $C.$ Since $\gamma_1+\gamma_2 \in C_{max}^{\bot}$ and $\textbf{1}\in C_{max}$ we have on the one hand $w(\gamma_1+\gamma_2) \equiv 0 \mod 2.$ On the other hand we have $w(\gamma_1+\gamma_2)=w(\gamma_1)+w(\gamma_2)-2\mu(\gamma_1,\gamma_2) \equiv 2 + w(\gamma_2) -2 \equiv w(\gamma_2) \mod 4.$ Thus either $C_1$ and $C_2$ are both singly-even or both doubly-even.
	$\hfill\square$

	\begin{theorem}\label{neighbor-weight}
		Let $C_1$ be a binary singly-even self-dual $(n,k,d)$-code and $C_2$ a binary doubly-even self-dual code. If $C_1$ and $C_2$ are neighbors, i.e.~they have the same maximal doubly-even subcode $C_{k-1},$ then all singly-even codewords in $C_{k-1}^{\bot}$ have weight $w,$ where $d\leq w \leq n-d.$
	\end{theorem}
	
	\textbf{Proof. } There exist $\gamma_1, \gamma_2\in C_{k-1}^{\bot}$ such that $C_1=\langle C_{k-1} \cup \{\gamma_1\}\rangle$ and $C_2=\langle C_{k-1} \cup \{\gamma_2\}\rangle.$ Thus $w(\gamma_1)\equiv 2 \mod 4$ and $w(\gamma_2)\equiv 0 \mod 4.$ Now we denote $S:=C_{k-1}^{\bot}\setminus (C_1\cup C_2).$ Then $\mid S\mid =2^{k-1}.$
		Then all codewords of $C_1\setminus C_{k-1}$ have the form $a+\gamma_1$ for some $a\in C_{k-1}$ and  all codewords of $C_2\setminus C_{k-1}$ have the form $a+\gamma_2$ for some $a\in C_{k-1}.$
		Further all codewords of $S$ have the form $a+\gamma_1 +\gamma_2$ for some $a\in C_{k-1}.$
		
		Since $C_1$ and $C_2$ are self-dual we have $\gamma_1 \not \perp \gamma_2.$ Thus $\mu(\gamma_1,\gamma_2)\equiv 1 \mod 2.$ Then
		
		\begin{equation}
			w(\gamma_1 + \gamma_2) = w(\gamma_1)+w(\gamma_2)-2\mu(\gamma_1,\gamma_2) \equiv 2 + 0 - 2 \equiv 0 \mod 4 
		\end{equation}
		
		Since $\gamma_1 + \gamma_2 \in C_{k-1}^{\bot}$ we have $\mu(a,\gamma_1 + \gamma_2) \equiv 0 \mod 2$  for all $a\in C_{k-1}.$ Thus 
		
		\begin{equation}
			w(a+\gamma_1 +\gamma_2) = w(a) + w(\gamma_1 + \gamma_2) -2\mu(a,\gamma_1 + \gamma_2) \equiv 0 + 0 - 0 \equiv 0 \mod 4
		\end{equation}
		
		which means that $S$ consists only of codewords with doubly-even weight.
		
		Since $\mid C_{k-1}^{\bot}\mid = 2^{k+1} = \mid S\mid + \mid C_2\mid + \mid C_1\setminus C_{k-1}\mid = 2^{k-1} + 2^k + 2^{k-1},$ the number of singly-even codewords in $C_{k-1}^{\bot}$ is $2^{k-1}.$ Therefore, for all singly-even codewords $c\in C_{k-1}^{\bot}$ implies $c\in C_1.$ 
	$\hfill\square$
	
	\begin{remark}
		Theorem \ref{neighbor-weight} can help us to make the search for singly-even $(n,k,d)$-codes faster, since it is a strong condition that in the dual code there cannot be any singly-even word with weight smaller than $d.$
	\end{remark}
	
	\section{Torch code searching package}\label{sec:4}
	
	Based on the statements of Section \ref{sec:2} and Section \ref{sec:3} we build up a computer-based code searching package.
	In this section we describe the software implementation used for searching for linear codes. The package, named Torch, is written in Python 3.x, and uses Sagemath 9.x (available at \textit{https://www.sagemath.org/}) for most linear code operations. Some operations use the GAP guava package (available at \textit{https://www.gap-system.org/Packages/guava.html}) or Magma (available at \newline \textit{http://magma.maths.usyd.edu.au/magma/}) when it is more efficient than the solutions provided by Sagemath. The source code repository is currently not publicly available.
	
	The goal of the package is to provide a reusable, modular solution for implementing linear code search algorithms, where adding a new search condition or a modified algorithm for a computation step can be done as quickly as possible. To achieve this, Torch uses a custom dependency injection class which reads a .json format configuration file, describing the relationships between the submodules, instantiates the necessary objects and assembles the search algorithm described in the configuration.
	
	The default configuration describes a depth-first search algorithm which builds a generator matrix row by row. When searching for an \textit{(n,k,d)} linear code, the algorithm starts with an $(n,1,d)$ matrix, and when the search is at depth $k'$, then the generator matrix being built has dimension $k'$. When the search reaches depth $k$, a solution is found. At each step of the search, a number of conditions are checked to see which possible new rows should the algorithm use to continue, and which branches of the search tree can be discarded. 
	
	Some of the modules in this configuration include:
	
	\begin{itemize}
		\item \textit{mu$\_$table}: creates a table of possible $\mu$ values between codewords, based on the target code's $(n,k,d)$ and type.
		\item \textit{starters}: generator function that returns the possible $(n,1,d)$ matrices that will be the root nodes of the search.
		\item \textit{descendants}: generator function that takes a $k'$ dimensional matrix, $k'<k$, as input, and returns all $(n,k,d)$ generator matrices which have the input matrix as their first $k'$ rows.
		\item \textit{children}: generator function that takes a $k'$ dimensional matrix, $k'<k$, as input, and returns all $(n,k'+1,d)$ generator matrices which have the input matrix as their first $k'$ rows. Can be given a set of conditions, which filter out possible solutions. Only those matrices are returned which pass all conditions.
		\item \textit{saver}: provides the option to save a linear code object to a file when a code with the target parameters are found. The default \textit{saver} uses the \textit{pickle} python package to serialize objects.
		\item \textit{logger}: provides the option for other modules to log their activity. The default logger logs to the console.
	\end{itemize} 
	
	The search is carried out by calling the \textit{descendants} module (which, in turn, recursively calls the \textit{children} module) on all the elements returned by the \textit{starter} module.
	
	All modules can be instantiated on their own, for testing or experimental purposes. Most modules have multiple implementing classes. In some cases, a factory pattern is used to decide between implementations during build time, based on information such as the target linear code's type. In other cases, modules can switch between implementations during runtime to try to pick the most efficient solution, based on criteria such as the current code's dimension and other properties. 
	
	At the top level, the Torch package includes a function named \textit{get$\_$code()}, which takes the parameters $(n,k,d)$ and code type. Other optional parameters include a custom configuration file and arbitrary values to be passed to submodules, such as silent mode, ascending/descending order, save file destination and many others. Once instantiated, \textit{get$\_$code()} acts as a generator function yielding Sagemath linear code objects of the given parameters. The default configuration is an exhaustive search, with conditions that ensure that at least one linear code from each permutation equivalence class is yielded.
	
	\section{Code search for $n\leq 30$}\label{sec:5}
	
	In Table~\ref{table:xyz}, we give some computation results which were achieved by our code searching program. We give the number of found codes. These codes are not necessarily inequivalent. In contrast, any existing binary $(n,k,d)$-code of Type I is equivalent to at least one of the found codes.
	\begin{table}[H]
		\centering
		\caption{Type I codes of small length.}\label{table:xyz}
		\begin{tabular}{|c|c|}
			\hline
			(n,k,d) & Number of found codes\\
			\hline
			(12,6,4) & 2 \\
			\hline
			(14,7,4) & 6 \\
			\hline
			(16,8,4) & 43 \\
			\hline
			(18,9,4) & 556 \\
			\hline
			(20,10,4) & 1977+ \\
			\hline
			(22,11,6) & 242 \\
			\hline
			(24,12,6) & 1279+ \\
			\hline
			(28,12,6) & 350+ \\
			\hline
			(30,15,6) & 6000+ \\
			\hline
			
		\end{tabular}
	\end{table}
	
	Where we write $+$ that means there are probably more codes than we found, because we stopped the code searching program at this point. The searching results depend not only on $n,k,d,$ but also on the searching strategy. For example, when searching self-dual $(30,15,6)$-codes we experienced that the searching program finds more than 6000 codes within one hour if we restrict the row weights to $10.$ Where the number of found codes is completely given, that means there exists no other $(n,k,d)$-code which is not equivalent to one of the found codes. On the other hand, the number of found codes is not equal to the number of inequivalent codes, but in some cases it is near to it (because of the hierarchy in the generator matrix in the code searching program).
	
	\section{Conclusion}\label{sec:6}
	
	In the current paper, we introduced a code searching program which is intended to use all good facts we know about Type I codes. Beside this, the code searching program is also able to generate Type II codes or any $(n,k,d)$ linear codes for relatively small $n.$ In Section \ref{sec:5} we give the number of found Type I codes for $12\leq n\leq 30.$ This means the code searching program found a generator matrix in standard form for each found code. Most of the codes were found within some minutes. Therefore the strength of this program lies in the capability to give generator matrices fast. Those generator matrices can be used further in cryptographic systems, where small code length is required, e.g.~for mobile applications.
	

	\printbibliography
	
\end{document}